\documentclass[prl,aps,showpacs,floatfix,byrevtex,amsmath,
superscriptaddress,amssymb,twocolumn]{revtex4}
\usepackage{graphicx}

\begin{document}

\preprint{PREPRINT}

\title{Overcharging of DNA in the presence of salt:\\
  Theory and Simulation }

\author{Markus Deserno}
\affiliation{Max-Planck-Institut f\"ur Polymerforschung
      Ackermannweg 10, 55128 Mainz, Germany}
\affiliation{Department of Chemistry and Biochemistry
      UCLA, 405 Hilgard Avenue, Los Angeles CA 90095, USA}

\author{Felipe Jim\'enez-\'Angeles}
\affiliation{Programa de Ingenier\'{\i}a Molecular, Instituto Mexicano del Petr\'oleo,
L\'azaro C\'ardenas 152, 07730 M\'exico, D. F., M\'exico}
\affiliation{Departamento de F\'{\i}sica, Universidad Aut\'onoma Metropoloitana-Iztapalapa,\\
Apartado Postal 55-334, 09340 D.F. M\'exico}

\author{Christian Holm}
\affiliation{Max-Planck-Institut f\"ur Polymerforschung
      Ackermannweg 10, 55128 Mainz, Germany}

\author{Marcelo Lozada-Cassou}
\affiliation{Programa de Ingenier\'{\i}a Molecular, Instituto Mexicano del Petr\'oleo,
L\'azaro C\'ardenas 152, 07730 M\'exico, D. F., M\'exico}
\affiliation{Departamento de F\'{\i}sica, Universidad Aut\'onoma Metropoloitana-Iztapalapa,\\
Apartado Postal 55-334, 09340 D.F. M\'exico}

\frenchspacing

\hyphenation{par-ti-cu-lar-ly ex-pe-ri-men-tal coun-ter-ion li-quids
  in-ter-mo-le-cu-lar ana-ly-ti-cal-ly theo-re-ti-cal theo-ri-es
  cor-res-pond}


\newcommand{\romB}{{\operatorname{B}}}
\newcommand{\romD}{{\operatorname{D}}}
\newcommand{\romHNC}{{\operatorname{HNC}}}
\newcommand{\romLJ}{{\operatorname{LJ}}}
\newcommand{\romLPB}{{\operatorname{LPB}}}
\newcommand{\romMD}{{\operatorname{MD}}}
\newcommand{\romPB}{{\operatorname{PB}}}

\newcommand{\romb}{{\operatorname{b}}}
\newcommand{\romca}{{\operatorname{ca}}}
\newcommand{\romcut}{{\operatorname{cut}}}
\newcommand{\romd}{{\operatorname{d}}}
\newcommand{\romr}{{\operatorname{r}}}
\newcommand{\romrod}{{\operatorname{rod}}}
\newcommand{\roms}{{\operatorname{s}}}

\newcommand{\Cmm}{{C/m$^2$}}

\newcommand{\etal}{{\em et al.\/}}
%
\newcommand{\ch}{}
\newcommand{\md}{}
\newcommand{\D}{\displaystyle}


\begin{abstract}
  A study of a model rod-like polyelectrolyte molecule immersed into a
  monovalent or divalent electrolyte is presented. Results for the local concentration profile,
  mean electrostatic potential, charge distribution function and $\zeta-$potential are obtained
  from hypernetted-chain/mean spherical approximation (HNC/MSA) theory
  and compared with molecular
  dynamics (MD) simulations.  As a particular case, the parameters of
  the polyelectrolyte molecule are mapped to those of a DNA molecule.
  Both, HNC/MSA and MD, predict the
  occurrence of overcharging, which is not present in the
  Poisson-Boltzmann theory. Futher an excellent qualitative, and in some cases quantitative, agreement
  between HNC/MSA and MD is found.  Oscillations observed in the mean electrostatic potential, local
  concentration profiles, as well as the curvature of the $\zeta$-potential 
  are discussed in terms of the observed overcharging effect. Particularly interesting results are
  a very non-monotonic behavior of the
  $\zeta$-potential, as a function of the rod charge density,
  and the overcharging by {\em monovalent} counterions.
\end {abstract}

\date{\today}

\maketitle

\section{Introduction}

``Polyelectrolytes are polymers bearing ionizable groups, which, in
polar solvents, can dissociate into charged polymer chains (macroions)
and small counterions'' \cite{barrat96a}. The combination of
macromolecular properties and long-range electrostatic interactions
results in an impressive variety of phenomena. It makes these systems
interesting from a fundamental as well as a technological point of
view.  A thorough understanding of polyelectrolytes has become
increasingly important in biochemistry and molecular biology. This is
due to the fact that virtually all proteins, as well as the DNA, are
polyelectrolytes.  Their interactions with each other and with the
charged cell-membrane are still far from being fully understood, which
is partly due to the intricate coupling between ion distribution and
chain conformation.

A first approach to the problem is to fix the conformation of the
chain and to focus on a detailed description of the counterion
distribution.  Usually polyelectrolytes stretch due to the
electrostatic repulsion of their charged groups. Moreover, many
important polyelectrolytes have a large intrinsic stiffness (e.g.,
DNA, actin filaments or microtubules).  Therefore, a rod-like
conformation is an obvious first choice. The remaining problem of
charged rods immersed into solution is much easier, but is still far
from being exactly solvable.

An additional approximation, which is frequently used in theoretical
descriptions, is to completely integrate out the counterionic degrees
of freedom. On a linearized mean-field level this yields a
Debye-H\"uckel-like theory characterized by a screened Coulomb
potential between charged monomers.  To obtain the correct physical
properties, one uses an {\it effective} Yukawa potential, which in
turn requires adjustable parameters such as an effective
polyelectrolyte radius and charge.  In this way the dependence of the
ionic structure, hereafter called electrical double layer (EDL),
around two or more polyelectrolytes, as a function of the
polyelectrolyte-polyelectrolyte distance, is not accounted for.  This
information, however, is most relevant for the understanding of
polyelectrolyte aggregation or self-assembling.  This effect is more
important for polyelectrolytes in low concentration added salt
solution, since for this case the ionic screening is weak. This has
been shown to be relevant for charged plates and charged spherical
macroions \cite{lozada90,sanchez92,jimenez00}.

A common further approximation assumes that the investigation of a
small sub-volume containing only one rod and its counterions will
suffice to unveil much of the interesting physics. The main
justification for this approach is that the sub-volume has zero net
charge. Moreover, the counterions will also efficiently screen higher
order multipoles. Hence, the interactions between two such
sub-volumes, which are neglected when focusing on just one rod, will
be fairly weak.  This approximation is called cylindrical cell model
and it provides the framework for our simulation calculation. 

{\md One attractive feature of the cell model is that in the salt-free
case the nonlinear PB equation can be solved {\em exactly} in this
geometry \cite{alfrey51,fuoss51}. It also displays in a clear fashion
the effect of partial counterion {\em condensation}
\cite{manning69a,oosawa71a}.  While a charged sphere loses all its
counterions upon dilution, a charged plane keeps all of them.  For a
charged cylinder the fraction of ions, which upon dilution remain in
the vicinity of the macroion, can be anywhere between $0$ and $100$\%.
Addition of salt increases the screening of the charged rod. If the
salt content is large enough, the electric field will have decayed to
zero before the cell radius is reached. It is then permissible to
extend the latter to infinity. This is the approach that we use for
our integral equation calculations.}

The PB theory shows its limitations for systems where {\ch ion}
correlations become important \cite{deserno00a}.  It cannot predict the
attractive forces which are seen experimentally in DNA solutions
\cite{bloomfield91a} and which have been also found in various
simulation \cite{valleau91,nilsson91a,lyubartsev98a} and integral
equation \cite{lozada86,kjellander86,lozada90b} studies in the
presence of multivalent counterions.

In this paper we present {\md studies} for a rod immersed in a salt
solution which PB theory fails to describe correctly, namely, the
possibility of overcharging a single rod, resulting in an effective
charge reversal, and the appearance of a non-monotonic zeta-potential
\cite{lozada83a,gonzales85a,kjellander97}. This effect is most
relevant in electrophoresis experiments \cite{hidalgo96}. The
overcharging of a macroion modifies in a non-linear fashion the
electrophoretic mobility as a function of the polyelectrolyte charge
and salt concentration. In the classical electrophoresis theory of
Wiersema, O'Brien and White \cite{wiersema66,obrien78}, which is a
linear theory based on the Poisson-Boltzmann (PB) description of the
EDL around the polyelectrolyte, this effect is not included.

The EDL obtained through the well established integral equation
approach does predict overcharging \cite{lozada83a,gonzales85a}. The
overcharging effect has been observed in Monte Carlo (MC) simulations
for planar \cite{valleau80}, cylindrical \cite{vlachy86} and spherical
\cite{degreve93,degreve95} geometries. {\md Moreover}, several years
ago, Gonzales-Tovar \etal \cite{gonzales85a} predicted that
for some macroion charge and electrolyte concentration the
overcharging produces an electrophoresis mobility reversal. This
mobility reversal has been experimentally observed
\cite{strauss54,elimelech90,dubin99,dubin00}. This effect has been
taken into account in the electrophoresis theory by Lozada-Cassou
\etal \cite{lozada99}. An excellent qualitative agreement with
experimental results \cite{hidalgo96,elimelech90,dubin00} is found.  This
overcharging effect and the predicted mobility reversal have been
recently addressed by Deserno \etal \cite{deserno00a} and
Shklovskii \cite{shklovskii99,shklovskii00}.

{\md In Ref.~\cite{deserno00a} ion distributions are studied via the
integrated charge distribution function, $P(r)$, and overcharging is
addressed based on MD simulations. In the present paper we extend this
research by approaching the overcharging effect through a well
established integral equation theory \cite{lozada83a,gonzales85a} as
well as more detailed molecular dynamics (MD) simulations.}  We
compute $P(r)$, the concentration profiles $g_i(r)$, the mean
electrostatic potential profile $\psi(r)$, and the
$\zeta$-potential. A comparison of the integral equation results with
those from MD is presented.


\section{Simulational details of the model system}\label{sec:model_system}

\subsection{Generating a cell-geometry}

\begin{figure}
\includegraphics[width=7.0cm]{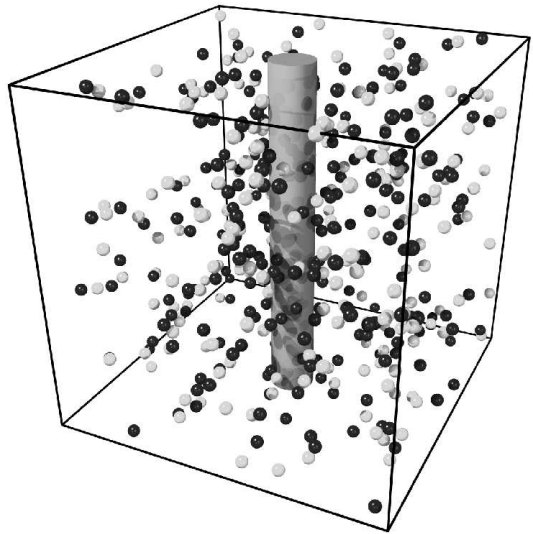}
  \caption{Realization of the cell model. A rod of length $L_\romb$
    placed parallel to an edge of a cube of side length $L_\romb$
    yields an infinite square array of infinitely long rods upon
    periodic replication of the cubic box. The dark particles are the
    counterions or positive salt ions and the bright particles are the
    negative salt ions. This particular snapshot contains 36 divalent
    counterions, 220 positive and 220 negative divalent salt
    ions.}\label{pic:cellpicture} \end{figure}

Compared to the spherical cell model, the cylindrical one presents one
additional but crucial complication: the charged rod is infinitely
long.  Several methods have been proposed in the literature to handle
this problem.  They essentially all use as a unit cell a hexagonal
prism with a certain height.  This approximates the cylindrical cell
by the ``most round'' space-filling object. In this publication we
take a computationally even simpler approach by using a {\em cubic\/}
unit box of side-length $L_\romb$ and placing the DNA parallel to one
of the edges (see Fig.~\ref{pic:cellpicture}). The justification for
this approach is as follows: In a high-salt environment the charged
rod will be screened within a distance much smaller than the size of
the box.  The potential will therefore become constant long before the
deviations from a cylindrical cell will be felt. It is thus largely
irrelevant, whether these deviations are of hexagonal or of cubic
type.  When comparing with the cylindrical cell model, a radius
$R=L_\romb/\sqrt{\pi}$ is appropriate, since it yields the same area
per rod.

The main advantage of this approach is that such a system can be
treated with the plain cubic Ewald sum or one of its mesh-upgrades
\cite{frenkel96a,deserno98a,deserno98b,hockney88a}.  This permits
a very efficient way for computing the long-range electrostatic
interactions.


\subsection{Interaction potentials and DNA mapping}

A specification of two interaction potentials is necessary to describe
the model system: {\em (i)\/} an excluded volume interaction preventing
two particles from occupying the same position in space and {\em
(ii)\/} the long-range Coulomb potential.

For the excluded volume interaction between the ions we use the
following potential:  
\begin{equation}
  V^*_{\text{ion-ion}}(r) \; = \; \left\{
\begin{array}{c@{\quad:\quad}l}
      4\epsilon\left[\left(\D\frac{a}{r}\right)^{12} \!-\;
        \left(\D\frac{a}{r}\right)^6 + \D\frac{1}{4}\right] & 0 < r \le
      r_\romcut\equiv 2^{1/6}a \\[2ex] 0 & r_\romcut < r.  
\end{array}
  \right.  \label{eq:shifted_LJ_potential} 
\end{equation} Without the
cutoff this would be the common Lennard-Jones potential describing
particles with a diameter $a$ and an attractive potential minimum of
$\epsilon$. In order to achieve pure repulsion, we cut the potential at
the minimum and shift it up such that it smoothly goes to zero there.
As a consequence, $\epsilon$ becomes largely irrelevant. Since two ions
a distance $a$ apart have the (repulsive) interaction energy
$\epsilon$, we make the natural choice $\epsilon=k_\romB T$.

In order to give the rod a larger radius than the ions, we employ an ion-rod
potential similar to Eq.~(\ref{eq:shifted_LJ_potential}), in which $r$ is
replaced by $r-r_\roms$. This shifts the hard core a distance $r_\roms$
towards larger radii and gives a distance of closest approach of $r_\roms+a$.
Since the ions have diameter $a$ this corresponds to a rod radius $r_0$ of
$r_0 = r_\roms+a/2$. Of course, in this case $r$ is a cylindrical and not a
spherical radial coordinate.

The electrostatic interaction energy between two ions with charge $z_i
e_0$ and $z_j e_0$, is given by 
\begin{equation}
  V^{\text{el}}_{i j}(r) \; = \; k_\romB T \, \frac{z_i z_j
  \ell_\romB}{r}
  \qquad\text{with}\qquad
  \ell_\romB :=
  \frac{e_0^2}{4\pi\varepsilon_0\varepsilon_\romr k_\romB T},
  \label{eq:electric} 
\end{equation} 
{\ch where} $z_i$ is the valence of the $i$ species and $e_0$ the
elemental charge.  By definition the {\em Bjerrum length\/}
$\ell_\romB$ is the distance at which two unit charges have
interaction energy $k_\romB T$.  With the vacuum dielectric constant
$\varepsilon_0=8.85 \times
10^{-12}\text{C}^2\text{m}^{-2}\text{N}^{-1}$ and the dielectric
coefficient $\varepsilon_\romr=78.5$ applying to water at room
temperature, one gets a Bjerrum length of approximately
$7.14\,\text{\AA}$. The line charge density $\lambda$ of the rod is
modeled by placing unit charges along the rod axis at the distance $b
= e_0/\lambda$. The number of charges along the rod per Bjerrum length
is an important dimensionless measure of the line charge density and
is often referred to as the Manning parameter: 
\begin{equation} 
  \xi \; = \; \frac{\lambda\ell_\romB}{e_0} \; = \; \frac{\ell_\romB}{b}.
\end{equation} 
Its relevance lies in the fact that for $\xi>1$ the phenomenon of
counterion ``condensation'' is observed \cite{manning69a}.

Within the periodic boundary conditions employed during the
simulations, the presence of such long-range interactions poses both
mathematical and technical difficulties. We use an efficient FFT
accelerated Ewald sum, the P$^3$M algorithm, which scales almost
linearly with the number of charges \cite{deserno98a,deserno98b,hockney88a}.

The final step consists of explicitly mapping the parameters to a DNA
system in aqueous solution. This affects ion diameter $a$, rod radius
$r_0$, Bjerrum length $\ell_\romB$ and line charge density $\lambda$.
Our choice is presented in Table~\ref{tab:DNA_param}.

We performed Molecular-Dynamics simulations of the rod-systems using a
Langevin thermostat in order to drive the system into the canonical
state \cite{grest86a}. The real-space and Fourier-space part of the
electrostatic energy was used to check for equilibration. The presented
observables originate from averaging over roughly 1500 independent
configurations.


\section{HNC/MSA theory} 

The integral equation formalism is a well established statistical
mechanical approach that has been shown to be reliable when it is
applied to simple microscopic models of inhomogeneous fluids
\cite{henderson92a,lozada92}. A particularly successful integral
equation theory for inhomogeneous, charged fluids is the so-called
hypernetted-chain/mean spherical approximation (HNC/MSA)
\cite{degreve93,lozada96,lozada97,lozada98}. The HNC/MSA theory for a
model charged rod, immersed in an electrolyte, has been derived in the
past \cite{lozada83a,gonzales85a}. In this paper, we solve the HNC/MSA
equation for an infinitely long, hard, charged cylinder of radius
$r_0$, with uniform line charge density $\lambda$. The rod is immersed
in a two-component restricted primitive model electrolyte (RPM), {\md
i.e.,} a fluid of charged hard spheres of diameter $a$ with a centered
point charge $z_i e_0$.

The fluid electroneutrality condition is 
\begin{equation}
 \sum_{m=1}^{2}z_m \rho_m =0, \label{eq:electroneutrality}
\end{equation} 
where $\rho_m$ is the bulk concentration of species $m$.  In order to
satisfy the necessary condition of zero electrical field at infinity,
the rod charge is compensated by the charge induced in the fluid,
$\lambda'$:
\begin{equation}
  \lambda'\equiv 2 \pi \int^{\infty}_{r_0+a/2}\rho_{el}(r)
  r\,d{r}=-\lambda, 
  \;\text{with}\;
  \rho_{el}(r) \equiv e_0 \sum_{m=1}^{2} z_m \rho_{m}(r).
  \label{eq:lambda} 
\end{equation} 
The local concentration profile of the species $m$ is denoted as
$\rho_m(r)$.  The solvent is taken as a dielectric continuum of
dielectric constant $\varepsilon_\romr$. {\md For simplicity, rod and
solvent are assumed to have the same dielectric constant, in order to
avoid complications with dielectric boundaries}.

{\md It has long been recognized in physics that particles and fields
are equivalent in the sense that both are defined through their
interaction potentials.}  This simple fact has been applied in the
past to derive, in a straightforward manner, inhomogeneous integral
equation theories from the Ornstein-Zernike (OZ) equation for
homogeneous fluids \cite{lozada81a,lozada92}.  The homogeneous
three-component OZ equation is
\begin{equation}
 h_{i j}({\bf r}_{21}) \; =
    \; c_{i j}({\bf r}_{21}) + \sum_{m=1}^{3} \rho_m
     \int h_{i m}({\bf r}_{23}) c_{m j}({\bf r}_{13}) \,d{v}_{3}.
 \label{eq:ozh} 
\end{equation} 
where $h_{i j}({\bf r}_{21})$ and $c_{i j}({\bf r}_{21})$ are the
total and direct correlation functions, between particle $2$ located
at ${\bf r}_{2}$ and particle $1$ at ${\bf r}_{1}$ of species $i$ and
$j$, respectively, ${\bf r}_{21} \equiv  {\bf r}_{2}-{\bf r}_{1}$.

{\md In consequence}, one can consider a particle of a fluid as a
source of an external field or an external field as a particle of the
fluid. {\md Applying this simple idea to Eqn.~(\ref{eq:ozh})}, one can
think of one of the species, say $\alpha$, as made of infinite
cylindrical rods and the remaining {\md two species as ions}.  In
$\alpha$'s infinite dilution limit the cylinders are uncorrelated.
Hence, letting $\rho_\alpha \to 0$, the Ornstein-Zernike equation for
an ionic solution next to a charged cylinder {\md reads}
\begin{equation} 
  h_{\alpha j}({\bf r}_{21}) \; =
    \; c_{\alpha j}({\bf r}_{21}) + \sum_{m=1}^{2} \rho_m
     \int h_{\alpha m}({\bf r}_{23}) c_{m j}({\bf r}_{13}) \,d{v}_{3}.
 \label{eq:oz}
\end{equation}

{\md In the past, several approximations (or ``closures'') for the
direct correlation function have been suggested. Two of them are:}
\begin{eqnarray} 
  \ln (g_{ij}({\bf r}_{21}))& = &-\beta V_{ij}({\bf r}_{21})+
  h_{i j}({\bf r}_{21})- c_{i j}({\bf r}_{21}) 
  \label{eq:chnc}\\
  c_{ij}({\bf r}_{21}) & = & -\beta V_{ij}({\bf r}_{21})
 \label{eq:cmsa}.
\end{eqnarray} 
Eqn.~(\ref{eq:chnc}) and Eqn.~(\ref{eq:cmsa}) are known as the
hypernetted chain (HNC) equation and the mean spherical approximation
(MSA), respectively; $V_{ij}({\bf r}_{21})$ is the direct interaction
potential between species $i$ and $j$, and $\beta \equiv 1/k_\romB T$.

If the HNC closure is used for the direct correlation function between
the rod particle and the $j$ species, equation~(\ref{eq:oz}) becomes
\begin{eqnarray}
  g_{\alpha j}({\bf r}_{21}) \; &=& \exp \left\{
  -\beta  V_{\alpha j}({\bf r}_{21}) \right. \\ \nonumber
     &+& \left. \sum_{m=1}^{2} \rho_m
     \int h_{\alpha m}({\bf r}_{23}) c_{m j}({\bf r}_{13}) d{v}_{3}
     \right\},
  \label{eq:hnc}
\end{eqnarray}
In this scheme, the two-particles correlation functions $h_{\alpha
j}({\bf r}_{21})$ and $c_{\alpha j}({\bf r}_{21})$ correspond to the
one-particle total and direct inhomogeneous correlation functions
$h_{j}({\bf r})$ and $c_{j}({\bf r})$ for species $j$ of a fluid under
the influence of an external field produced by a rod particle. The
local concentration for {\md the} $j$ species is given by $\rho_j({\bf
r}) = \rho_j g_j({\bf r})$, where $g_j({\bf r}) = h_j({\bf r})+1$ is
called the reduced concentration profile. Therefore, the charge
concentration profile in Eqn.~(\ref{eq:lambda}) is given by
\begin{equation}
  \rho_{el}(r) \;= \sum_{m=1}^{2} z_m e_0 \rho_m g_{m}(r).
  \label{eq:rho_chrg}
\end{equation}

For this model the direct interaction potential between the rod and
the $j$ species of the fluid, $V_{j}(r)$, can be separated into two
parts: the hard sphere-hard rod term $V^*_{j}(r)$ and the electrostatic
interaction potential $V^{\text{el}}_{j}(r)$. The first takes into
account the fact that ions cannot penetrate or deform the cylinder
\begin{equation}
  V^*_{j}(r) \; = \; \left\{
    \begin{array}{c@{\quad:\quad}l}
      \infty   & 0 < r \le r_0+a/2 \\[2ex] 
             0 & r > r_0+a/2 
    \end{array}
   \right.  
  \label{eq:hard_term}
\end{equation} 
{\md The second can be found from Gauss' law and is given by}
\begin{equation}
  - \beta V^{\text{el}}_{j}(r) = 2 z_j {\xi} \ln(r) \qquad (r > r_0).
\end{equation}

In the MSA closure the homogeneous direct correlation function for a
RPM electrolyte has an analytical expression. This function can be
written as
\begin{equation}
  c_{m j}(s) \; =-\beta V^{\text{el}}_{{m j}}(s)+
    z_m z_j c^{\text{sr}}_{\text d}(s)+ c^{\text{hs}}_{\text s}(s).
  \label{eq:msa}
\end{equation}
In the first term appears the direct electric interaction potential
between the species of the fluid, which is given by
Eqn.~(\ref{eq:electric}), the second term is an electrical short range
function and the third is the direct correlation function for a hard
sphere fluid. When the MSA closure is used in the integral of
Eqn.~(\ref{eq:hnc}), one obtains the HNC/MSA equation for an ionic
fluid next to a charged rod.  Taking advantage of the cylindrical
geometry and {\md the fact that} the direct correlation function
between the ions depends only on their relative distance $s \equiv
|{\bf r}_{1}-{\bf r}_{3}|$, {\md Eqn.~(\ref{eq:hnc}) can after some
algebra} be written as
\cite{lozada83a,gonzales85a}
\begin{equation}
  g_j(r) \; = \exp \; \left\{ -\beta(z_j e_0 \psi (r)  + J_j(r)) \;
  \right\}, 
  \label{eq:hnc2}
\end{equation}
where $\psi(r)$ is the mean electrostatic potential
\begin{eqnarray}
  -\beta e_0 \psi(r) &=&  2 {\xi} \ln(r) \\ \nonumber 
    &+&2 \pi \ell_\romB \int^{\infty}_{r_0+a/2} \rho_{\text{cd}}(y)
   \ln((r^2+y^2+|r^2-y^2|)/2)y \romd y.  \label{eq:meanelectric}
\end{eqnarray}
The $J_j(r)$ terms are integrals of the short range terms of the direct
correlation function
\begin{eqnarray}
  - \beta J_j(r)  &=& \rho A(r) +z_j \int^{\infty}_{r_0+a/2} \rho_{\text{cd}}(y)
  L(r,y) \romd{y}\\
   &+& \int^{\infty}_{r_0+a/2} \rho_{\text{cs}} K(r,y)  \romd{y}
    ,
  \label{eq:kernel} 
\end{eqnarray}
with
\begin{eqnarray}
    \rho                & = & \sum_{m=1}^{2} \rho_m, \nonumber \\
    \rho_{\text{cs}}(y) & = & \sum_{m=1}^{2} \rho_m h_m(y), \nonumber \\ 
    \rho_{\text{cd}}(y) & = & \sum_{m=1}^{2} z_m\rho_m h_m(y), \nonumber
\end{eqnarray} 
and 
\begin{eqnarray}
  L(r,y)& = & 4y \int^{\phi_{max}}_0\,\romd{\phi}
    \int^{z_{max}}_0c^{\text{sr}}_{\text d}(s)\,\romd{z} \nonumber \\
  K(r,y)& = & 4y \int^{\phi_{max}}_0\,\romd{\phi}
    \int^{z_{max}}_0c^{\text{hs}}_{\text s}(s)\,\romd{z} \nonumber \\
  A(r)& = & -\int^{r_0+a/2}_{0} K(r,y) \,\romd{y} \nonumber 
\end{eqnarray} 
where $s^2=z^2+r^2+y^2-2ry \cos \phi$. 
In the {\md limit $a \to 0$ of point ions}, $c^{\text{sr}}_{\text
d}(s)$ and $c^{\text{hs}}_{\text s}(s) \to 0 $. Thus $J_j(r) \to 0$
\cite{lozada83a} and Eqn.~(\ref{eq:hnc2}) becomes
\begin{equation}
  g_m(r) \; = \exp \; \left\{ -\beta z_m e_0 \psi(r) \; \right\},
  \label{eq:pb1} 
\end{equation} 
where $\psi(r)$ is given by Eqn.~(\ref{eq:meanelectric}), which is the
solution of the well known PB differential equation for point ions
around a charged cylindrical electrode, i.e.,
\begin{equation}
  \frac{1}{r} \frac{\romd\psi(r)}{\romd r}+ \frac{\romd^2\psi(r)}{\romd r^2} \;=
  -\frac{e_0}{\varepsilon_0\varepsilon_\romr} \sum_{m=1}^{2} z_m \rho_m
  \exp \left\{ -\beta z_m e_0 \psi(r) \; \right\}.  
\end{equation}
Therefore, the integral equation version of the Poisson-Boltzmann
differential equation is \cite{lozada83a}
\begin{eqnarray}
 g_j(r) = \exp \left\{ 2 z_j {\xi} \ln(r) \right.\\
        +      \left. 2 z_j \pi \ell_\romB \int_{r_0+a/2}^{\infty} \rho_{\text{cd}}(y)
     \ln((r^2+y^2+|r^2-y^2|)/2)y \romd y \right\},
  \label{eq:pb2}
\end{eqnarray} 
which is also the HNC/MSA equation in the point ions limit. The
differences between these two theories are due to the ionic size
correlations which are partially taken into account in HNC/MSA theory
{\md but} which are ignored in the PB equation. In the past, it has
been shown that the size effects become important for strong field
interaction, poly-valent ions and high salt concentration
\cite{gonzales85a}.

The HNC/MSA and PB integral equations are numerically solved with
efficient finite element methods \cite{lozada90b,mier89}.  The
solution of Eqn.~(\ref{eq:hnc2}) takes one minute in a R12000
processor of an SGI machine. In our theoretical calculation we use the
same input parameters as in our MD calculations. The salt
concentration is obtained through
\begin{equation} 
  \rho_{\roms}=N_{\roms}/L_\romb^3, 
\end{equation}
where $N_\roms$ is the number of salt molecules used in the
simulation.  Nevertheless, we have to point out the differences in the
short range interaction used in both models. {\md While in HNC/MSA we
use hard particles, the MD for technical reasons uses the potential
from eq.~\ref{eq:shifted_LJ_potential}, which has some surface
softness.} This results in an effective excluded volume, whose actual
value depends on the interaction strength. At an interaction energy of
$k_\romB T$ the particle diameter is $a$. However, if the
particles attract each other strongly, they can come closer to each
other.

\section{Results}\label{ssec:PrDNA}

\begin{table}
    \small \begin{tabular}{lccc}\hline \hline
      parameter & symbol & value & value in LJ units \\ \hline \hline
      ion diameter & $a$ & 4.25 \AA & $a$ \\ ion valence & $z$
      & 2 & 2 \\ rod radius & $r_0$ & 7.86 \AA & $1.85\;a$ \\ line
      charge density (DNA) & $\lambda$ & $-e_0\big/1.7\,\text{\AA}$ &
      $-2.5\;e_0/a$ \\ Bjerrum length (water) & $\ell_\romB$ & 7.14 \AA
      & $1.68\;a$ \\ Manning parameter & $\xi$ & 4.2 & 4.2 \\ box size
      & $L_\romb$ & 122.4 \AA & $28.8\,a$ \\ corresponding cell radius
      & $R$ & 69.1 \AA & $16.2\;a$ \\ temperature & $T$ & 298 K &
      $\epsilon/k_\romB$ \\ \hline \end{tabular} 
  \caption{System parameters for the DNA
  simulations.}\label{tab:DNA_param} \end{table}

\begin{figure}
\includegraphics[width=7.0cm]{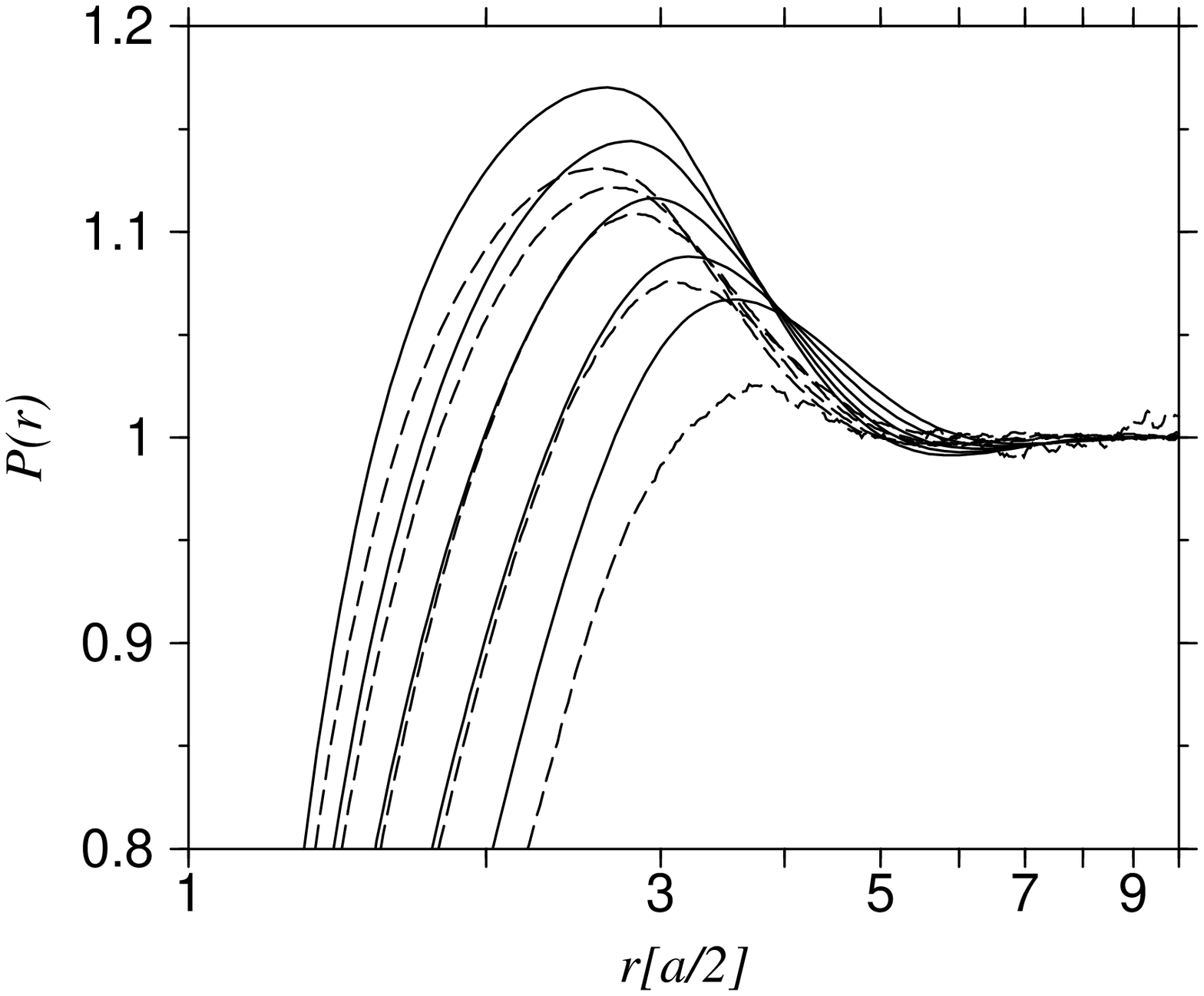}
\includegraphics[width=7.0cm]{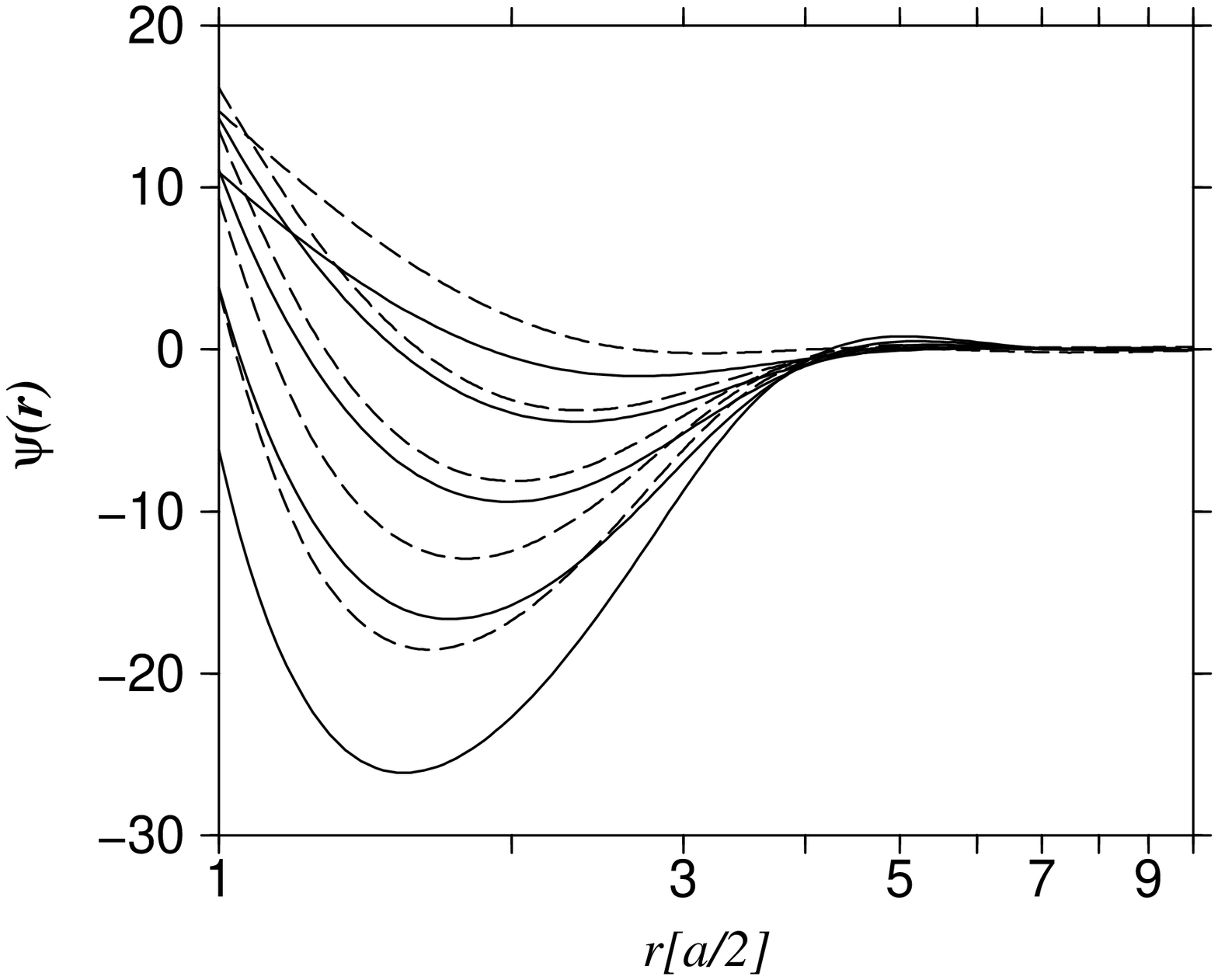}
    \caption{(a)Charge distribution function $P(r)$ 
    and (b) mean electrostatic potential $\psi(r)$ for the
    parameters from Tab.~\ref{tab:DNA_param}. The electrolyte is a 2:2
    salt of concentration 0.49 M. The distance $r$ is measured with
    respect to the cylinder surface in units of the ionic radius.  The
    solid and dashed lines correspond to the HNC/MSA and MD
    calculations, respectively. The different curves are for different
    values of the Manning parameter, $\xi = 2.1, 4.2, 6.3, 8.4, 10.5$,
    which correspond to surface charge densities of
    $\sigma=0.095,0.190,0.286,0.381,0.476$ \Cmm, respectively.  In
    $P(r)$, the surface charge density increases from bottom to top,
    whereas in the mean electrostatic potential it increases from top
    to bottom. The value $\xi = 4.2$ ($\sigma=0.190$ \Cmm)
    corresponds to DNA.}\label{fig:DNA_Pr_MEP_lambda}
\end{figure}

We will discuss the results in terms of the mean electrostatic potential,
$\psi(r)$, the local concentration profiles, $g_{m}(r)$, and the integrated
charge distribution function
\begin{equation}
  P(r) = \frac{1}{|\lambda|}
  \int_{r_0}^r \romd \bar{r} \; 2\pi \bar{r} \, \rho_{el}(\bar{r}).
\end{equation}
Figure~\ref{fig:DNA_Pr_MEP_lambda} shows the HNC/MSA and MD results
for $P(r)$ and $\psi(r)$, for five rod-systems, for which the line
charge density, and thereby the Manning parameter, has been
successively increased.  One system has a smaller Manning parameter
than DNA and three a larger.  All systems have a Debye screening
length which is much smaller than the Manning radius of the
corresponding salt free system.  According to the observations of Ref.
\cite{deserno00a} this means that the concept of Manning ``condensation''
is no longer meaningful. The rod charge is compensated by salt
screening.
%
%
%
According to simple PB theory, no remarkable features are to be
expected in the bulk solution.  Contrary to that, however, in
Fig.~{\ref{fig:DNA_Pr_MEP_lambda}a} $P(r)$ shows overcharging, i.e., the
rod charge is over-compensated at a certain distance from the rod.
For a given salt concentration 
the degree of overcharging is increased as the charge density increases. No overcharging saturation point is observed, 
at least for physical charge densities. While 
in Fig. {\ref{fig:DNA_Pr_MEP_lambda}a} overcharging is present even for a very low charge density, such as
 $\sigma=0.095$C/m$^2$  our HNC/MSA calculations show
overcharging for even lower rod charge densities. We will come back to this point later.
In Fig. {\ref{fig:DNA_Pr_MEP_lambda}a}, the maximum of overcharging is closer to the rod's suface as $\sigma$ 
increases. This effect has implication for the location of the so-called $\zeta-$potential, in electrophoresis 
experiments \cite{lozada99}. 
As pointed out before \cite{gonzales85a,valleau80}, overcharging implies
a change of direction in the local electrical field. Our MD and
HNC/MSA results are consistent with this fact, as shown in
Fig.~{\ref{fig:DNA_Pr_MEP_lambda}b}, where a minimum in
$\psi(r)$ is observed.
The value of $r$ at the  minimum of $\psi(r)$, i.e., where the electrical field is zero, 
concides with that at which $P(r)=1$, as it should.

Integral equation theories give much more reliable results than
Poisson-Boltzmann at high salt concentrations, poly-valent ions and/or
for fluids under the influence of strong external fields. This is
because they partially take into account correlations due to {\ch the}
ionic size, which PB theory ignores. It can be seen that the HNC/MSA
theory indeed {\ch correctly} predicts the occurrence of overcharging,
although it overestimates {\md its amount}. The qualitative agreement
between the HNC/MSA and MD results is excellent.  The quantitative
agreement is fair and particularly good for the DNA calculation.  In
the next section we will discuss the disagreement seen between the
HNC/MSA and MD results, for high and low cylinder charge densities.

\begin{figure}
\includegraphics[width=7.0cm]{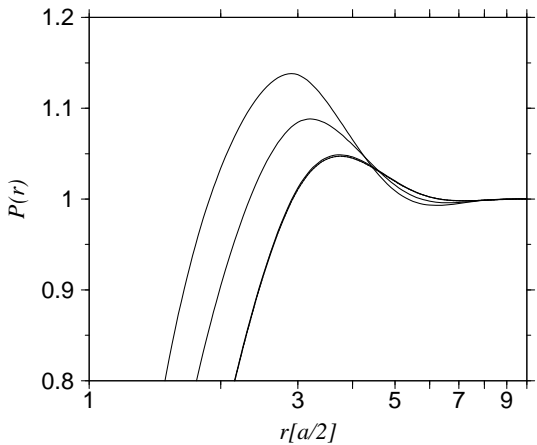}
  \caption{Charge distribution function from HNC/MSA, for a
    RPM electrolyte next to a charged cylinder. The distance $r$ is
    measured with respect to the cylinder surface in units of the
    ionic radius.  The electrolyte is a 2:2, 0.49 M salt. The results
    correspond to different cylinder radii, whereas the Manning
    parameter has been held constant at $\xi=4.2$. From top to bottom
    the cylinder radii are $r_0=3.86, 7.86, 110, 1200$ {\AA}, which
    correspond to the surface charge densities $\sigma= 0.388, 0.190,
    10^{-2}, 10^{-3}$ \Cmm, respectively. The curves for $r_0=110
    \text{\AA}$ and $r_0=1200 \text{\AA}$ are indistinguishable on
    this scale.}\label{fig:radius}
\end{figure}

For a given constant value of the Manning parameter, cylindrical
macroions with different diameters will produce different EDL
structures. In Figure~\ref{fig:radius}, HNC/MSA results for rods with
the same value of $\xi$ but different diameters {\md show} different
charge distributions $P(r)$.  This illustrates the simple fact that
the Manning parameter $\xi$ alone is not {\md sufficient} to specify
the ionic structure around cylindrical macroions. Since the ionic
structure depends directly on the field strength, it is often better
to use the cylinder's surface charge density $\sigma$, which is a
direct measure of the electrical field that penetrates into in the
solution. {\md Note, however, that $\xi$ and $\sigma$ are simply
related by $\xi = 2\pi r_0 \sigma \ell_\romB / e_0$.}

\begin{figure} 
\includegraphics[width=7.0cm]{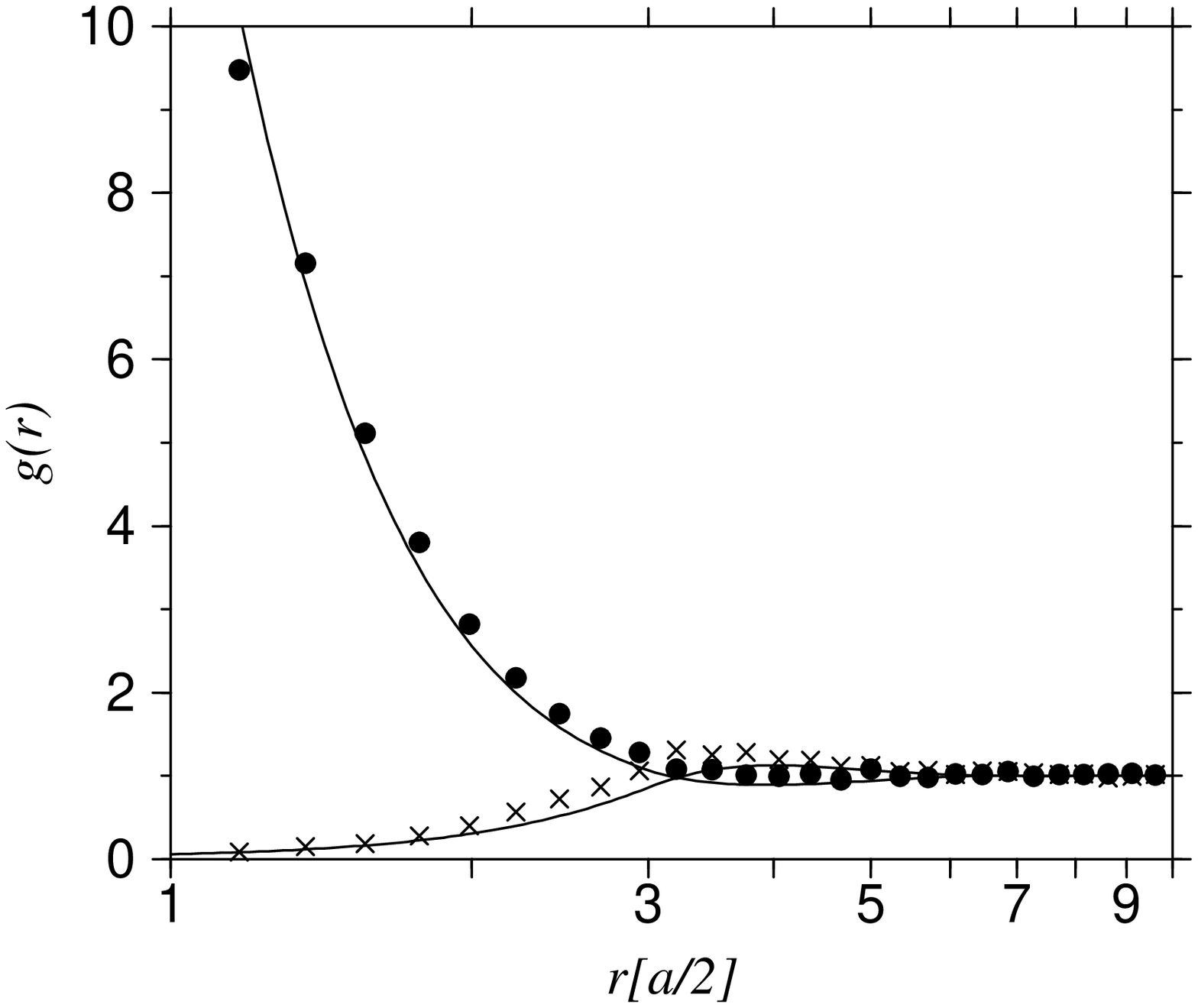}
\includegraphics[width=7.0cm]{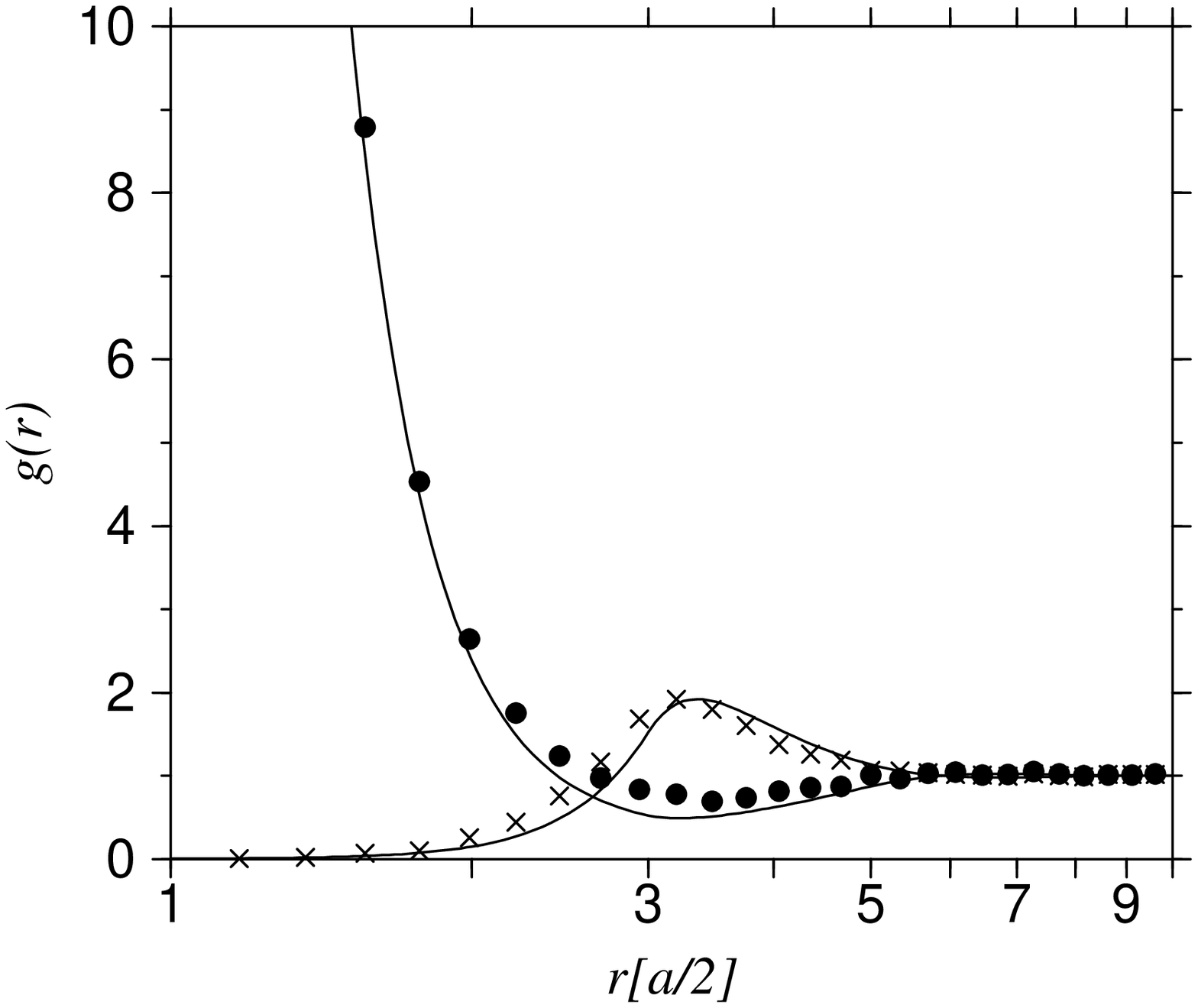}
  \caption{Reduced concentration profiles $g_{i}(r)$ for
    an electrolyte next to a charged cylinder. The electrolyte is a
    2:2, 0.49 M salt. The distance $r$ is measured with respect to the
    cylinder surface in units of the ionic radius. The system
    parameters are given in Tab.~\ref{tab:DNA_param}. Fig. {\ref{g21}a} shows the results
    corresponding to $\xi=4.2$ ($\sigma=0.190$ 
    \Cmm) and Fig. {\ref{g21}b} the results for $\xi=10.5$
    ($\sigma=0.49$ \Cmm).  The dots and crosses are the
    MD distribution functions for counterions and coions, respectively.
    The solid lines are the results from HNC/MSA.}
    \label{g21}
\end{figure}

In Figs. \ref{g21}a and  \ref{g21}b we show HNC/MSA and MD reduced
concentration {\ch profiles} for $\sigma=0.190$ \Cmm\ ($\xi=4.2$)  and $\sigma=0.49$ \Cmm\ ( $\xi=10.5$), respectively,
whereas all the other parameters are
the same as in Fig.~\ref{fig:DNA_Pr_MEP_lambda}. 
The value of $r$ at the  maximum of $P(r)$
matches with that where $g_{+}(r)=g_{-}(r)$, at least for symetric electrolytes. Therefore the later oscillation 
in the local concentration profiles, show the attraction (repulsion) to coions (counterions),
 due to a change in direction of the
effective electrical field. On the other hand, concentration profile oscillations are a consequence of the ionic
size correlations. Hence, overcharging results of electrostatic attraction and size correlations. 

\begin{table}
    \begin{tabular}{lcccc}\hline \hline
      {$\xi$} &
      {$\zeta_\romLPB$ [mV]} &
      {$\zeta_\romPB$ [mV]} &
      {$\zeta_\romHNC$ [mV]} &
      {$\zeta_\romMD$ [mV]} \\ \hline \hline0    &   0 &
      0    &  0     &  0     \\ 1.05 &  10.52 & 10.28 &  6.262 & 9.116
      \\ 2.1  &  21.04 & 19.48 & 10.99  & 14.62  \\ 4.2  &  42.09 &
      33.32 & 14.27  & 16.25  \\ 6.3  &  63.13 & 42.96 & 11.08  &
      13.80  \\ 8.4  &  84.17 & 50.16 &  3.771 &  9.263 \\ 10.5  &
      105.2  & 56.03 & -6.219 &  3.675 \\  \hline \end{tabular}
  \caption{$\zeta$-potential of a DNA-sized rod (see
  Tab.~\ref{tab:DNA_param})
    immersed into a 0.49 M electrolyte of 2:2 salt as a function of
    its Manning parameter $\xi$. The four predictions are from linear
    PB theory, PB theory, hypernetted-chain theory and MD simulation.
    The error in the latter is estimated to be of the order of 2\%.
    Figure~\ref{fig:zeta} visualizes the data.}\label{tab:DNA_zeta}
\end{table}

The first maximum seen in the coion reduced concentration profile implies a coion
concentration above its bulk value and indicates that the local electric field
is attractive to coions.
It is also observed that HNC/MSA overestimates the
contact value of the distribution function with respect to molecular dynamics
predictions. The overestimation of HNC/MSA can be associated with two facts:
{\em (i)\/} the difference in excluded volume used in both models; and 
{\em (ii)\/} HNC/MSA theory does not take into account all the size and charge
correlations.

\begin{figure}
\includegraphics[width=7.0cm]{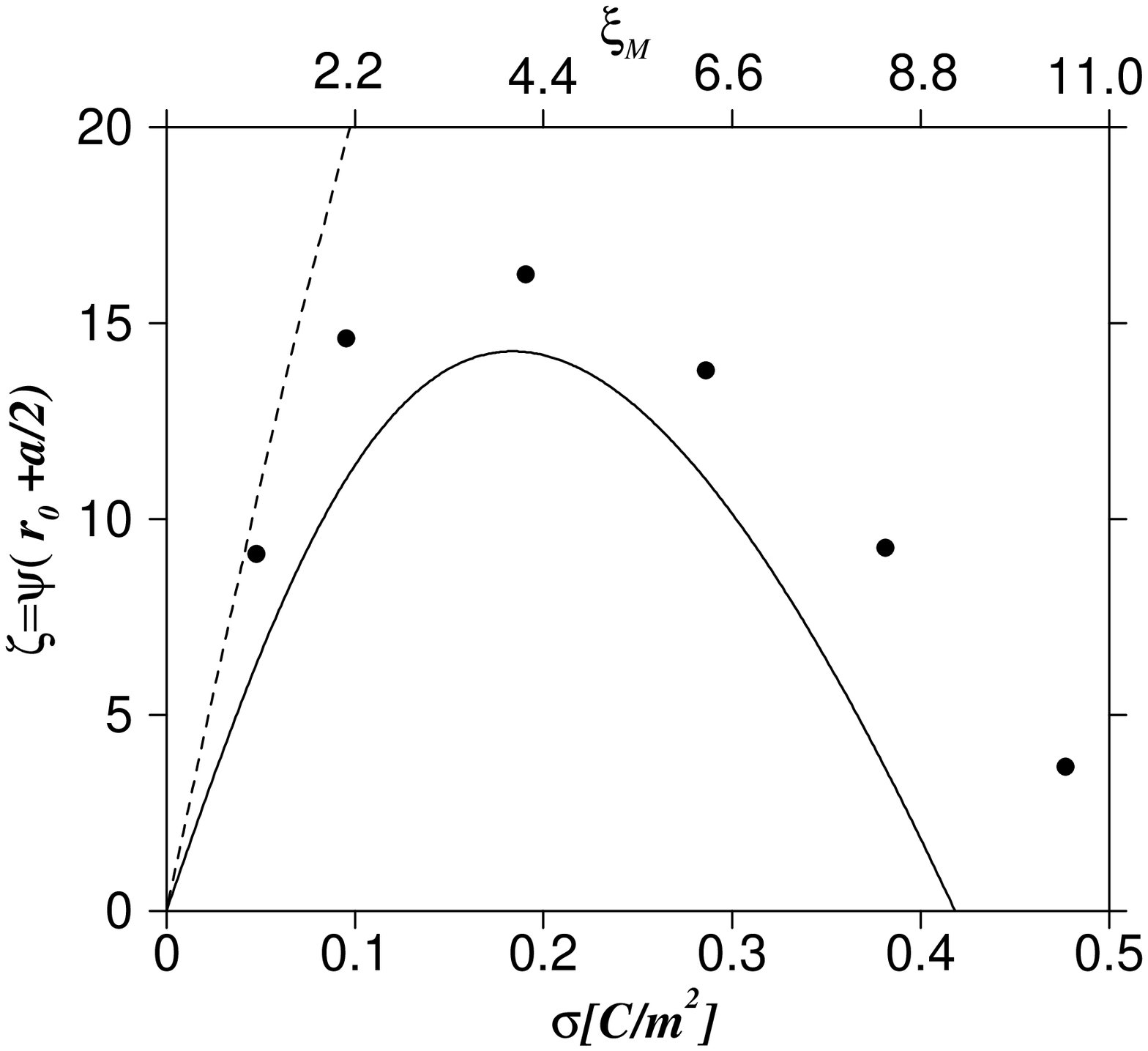}
  \caption{Zeta potential $\zeta\equiv\psi(r_0+a/2)$ as a function
    of surface charge density $\sigma$, for a cylinder immersed in a
    $0.49$ M, 2:2 electrolyte. The model parameters are given in
    Tab.~\ref{tab:DNA_param}.  The dotted line is the prediction of PB
    theory, the dots are the MD results and the solid line is the
    results from HNC/MSA calculations.}\label{fig:zeta}
\end{figure}


\begin{figure}
\includegraphics[width=7.0cm]{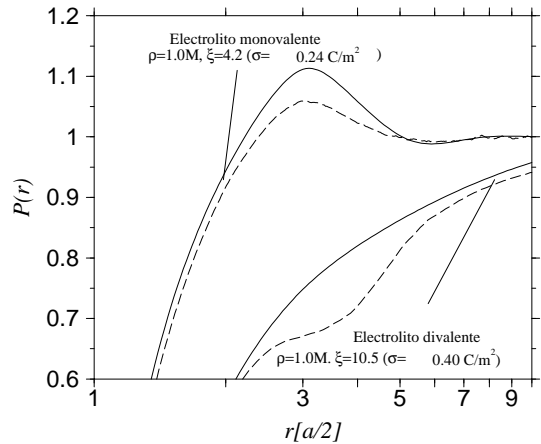}
  \caption{{\ch Charge distribution function from HNC/MSA (solid lines) and MD
      (dotted lines), for a RPM electrolyte next to a charged cylinder to
      which the ions have a distance of closest approach of $9.98 \text
      {\AA}$.  The distance $r$ is measured with respect to the cylinder
      surface in units of the ionic radius. In this plot a comparison between
      a low and a high excluded volume systems is presented. The system where
      overcharging occurs corresponds to a monovalent electrolyte with
      $\rho=1.0$ M, $\xi = 4.2$ ($\sigma = 0.240$ \Cmm), and large ions 
      ($a=7.4$\AA). The system where
      overcharging is not observed corresponds to a divalent electrolyte with
      $\rho=0.5$M, $\xi = 10.5$ ($\sigma = 0.40$ \Cmm), and small 
      ions ($a=1.0$\AA).}}
    \label{fig:energy-entropy}
\end{figure}


\begin{figure}
\includegraphics[width=7.0cm]{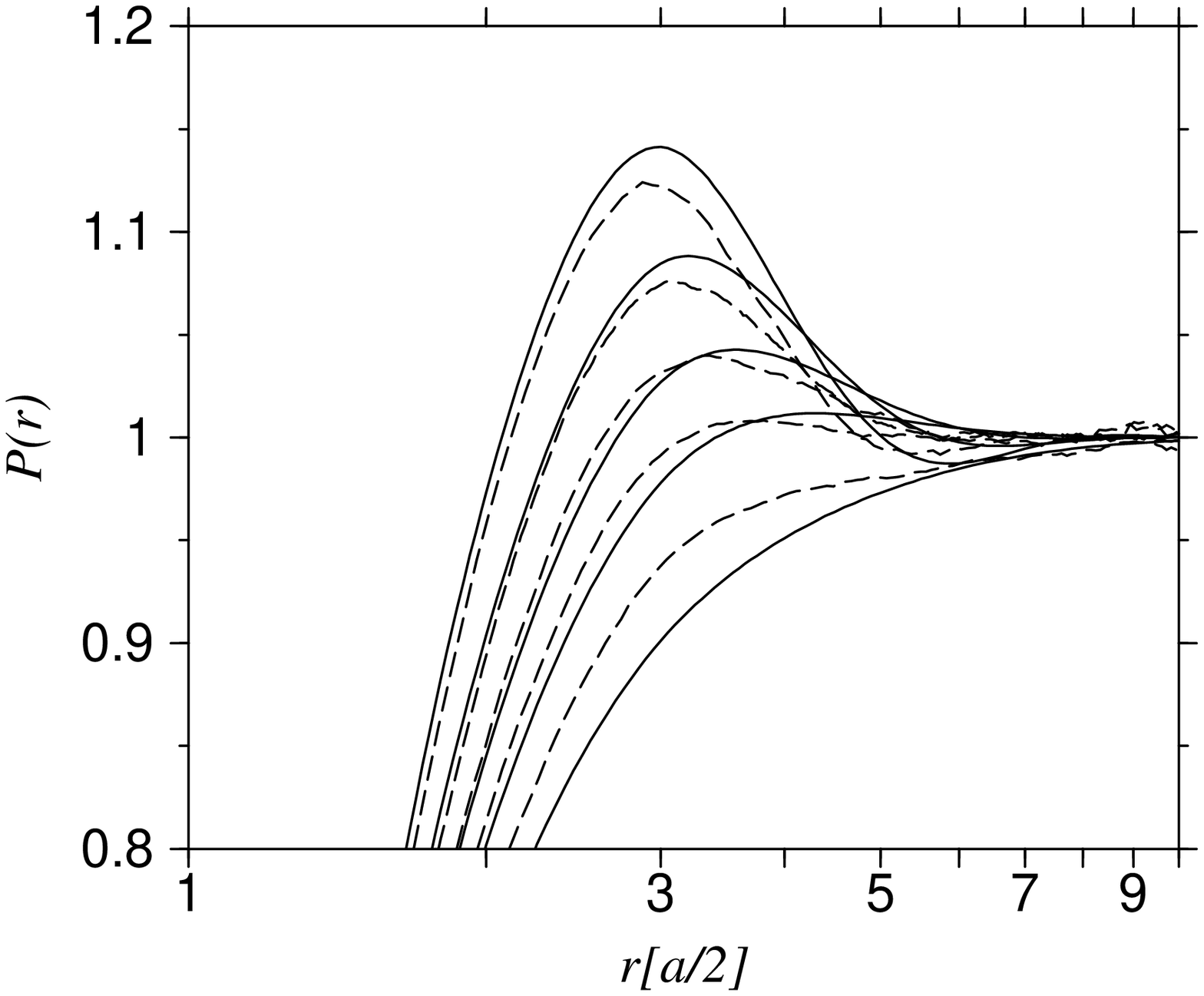}
\includegraphics[width=7.0cm]{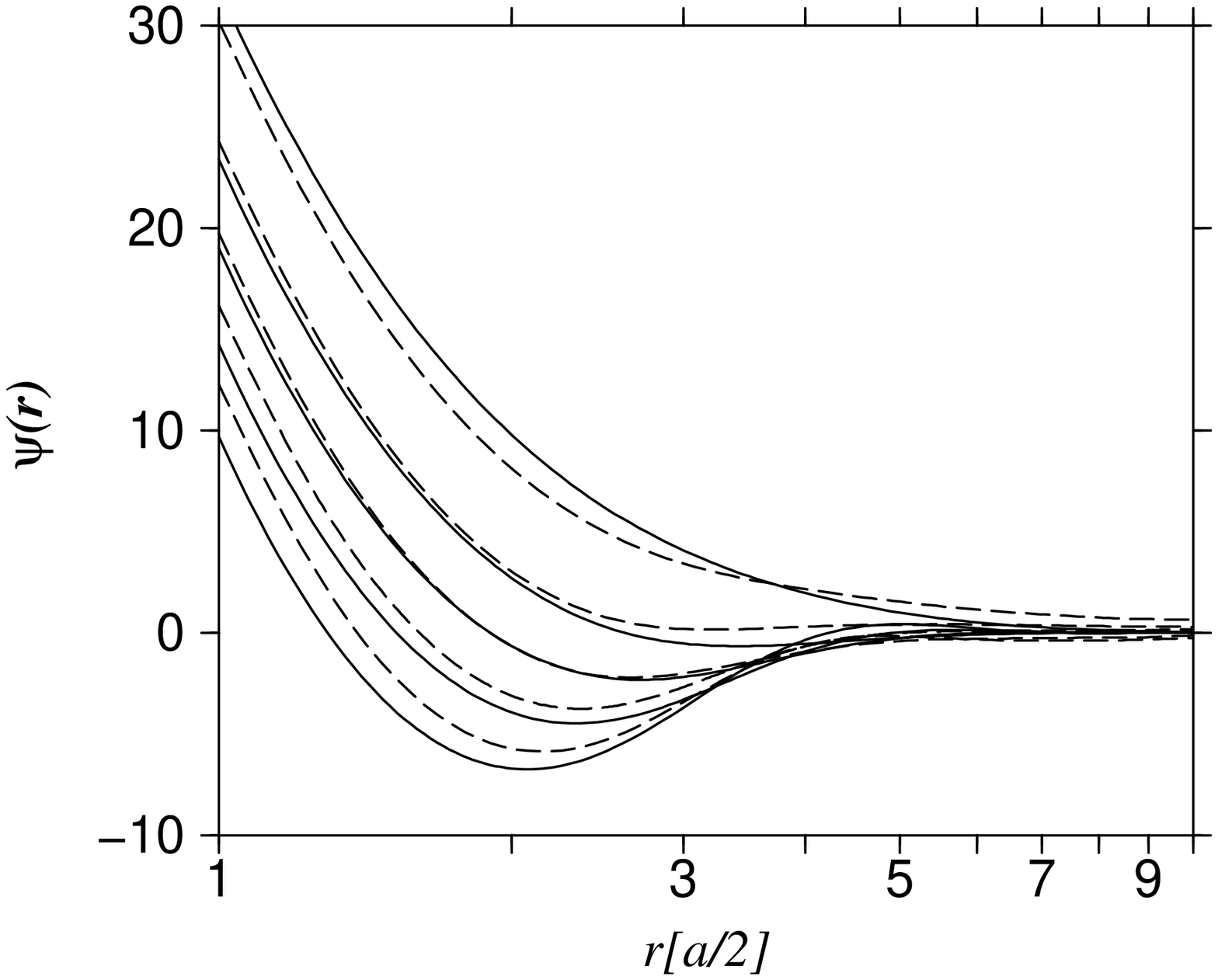}
    \caption{Charge distribution function $P(r)$ (left) and mean
      electrostatic potential $\psi(r)$ (right) results, for the system from
      Tab.~\ref{tab:DNA_param} with Manning parameter $\xi=4.2$. The distance
      $r$ is measured with respect to the cylinder surface in units of the
      ionic radius. The line styles have the same meaning as in
      Fig.~\ref{fig:DNA_Pr_MEP_lambda}.  The curves correspond to different
      salt concentration, $\rho_\roms = 0.12, 0.24, 0.34, 0.49, 0.68 $ M.  In
      the charge distribution function the salt content increases from bottom
      to top, in the mean electrostatic potential it increases from top to
      bottom.}
      \label{fig:DNA_Pr_MEP_rho}
\end{figure}

Another important phenomenon is indicated by the behavior of the mean
electrostatic potential in
{Fig.~\ref{fig:DNA_Pr_MEP_lambda}b}. It concerns the value of this
potential at the distance of closest approach between ions and the
rod,
\begin{equation}
  \zeta \; = \; \psi(r_0+a/2),
  \label{eq:zeta_potential} 
\end{equation}
which can be identified with the zeta-potential of electrophoresis
{\md and which is} of importance in the computation of electrophoretic
mobilities \cite{lozada99,hunter87a}.  In the PB theory for a charged
rod immersed into an ionic solution, and its linearized version,
$\zeta$ depends monotonically on $\xi$ \cite{gonzales85a}.  However,
Fig.~\ref{fig:DNA_Pr_MEP_lambda} shows that, in the presence of
overcharging, $\psi(r)$ can {\md become negative}.  This calls the
monotonic relation between $\zeta$ and $\xi$ into question.
Table~\ref{tab:DNA_zeta} summarizes various predictions for the
$\zeta$-potential as a function of Manning parameter together with
results from the molecular dynamics simulations. A graphical
illustration is given in Fig.~\ref{fig:zeta}.

Indeed, $\zeta$ is found to first increase with $\xi$, but from a certain
value on it decreases and would finally even become negative.  Note that this
would reverse the drift direction in electrophoresis measurements, as first
predicted by Gonzales-Tovar \etal \cite{gonzales85a} and recently
demonstrated, for spherical macroions, by Lozada-Cassou \etal \cite{lozada99}.
While nonlinear and linearized PB theory coincide with the data and with each
other for small Manning parameter, they completely fail to predict the
back-bending, which already sets in at comparatively small values of $\xi$.
The HNC/MSA theory captures this effect, but it underestimates the value of
the potential. However, comparisons for the $\zeta$-potential, as a function
of $\sigma$, between HNC/MSA and MC simulations for planar and spherical hard,
charged macroions, immersed into a RPM electrolyte, have been made in the past
\cite{degreve93,lozada98,lozada82}. For these two geometries, for 1:1
electrolytes and for 2:2, at around 0.5 M, as in our Fig.~\ref{fig:zeta}, the
HNC/MSA-MC agreement is excellent.  {\ch The disagreement seen in
  Fig.~\ref{fig:zeta} can be related to the different short-range potentials
  used in our MD and HNC/MSA calculations, which gives for the MD not a {\md
    sharply} defined ion-cylinder contact. This is important, since the
  $\zeta$-potential is evaluated exactly there.}
%
For planar and spherical macroions, PB $\zeta$ vs. charge curves are
{\ch also} above the HNC/MSA and MC curves, {\ch which is consistent
with our present observations.}

For a given rod charge and diameter, for any electrolyte solution
parameters, a higher $\zeta$-potential always implies a
lower counterions adsorption \cite{gonzales85a}.
%
A soft-repulsive short-range potential, as in our MD calculations,
allows the counterions to get closer to the rod.  This produces a
higher $\zeta$-potential in the MD simulation when compared to the
HNC/MSA results.  In the case of the PB theory, the neglect of the
ionic-size correlations allows a higher density of counterions next to
the rod, which in turn can never lead to overcharging.

\begin{figure}
\includegraphics[width=7.0cm]{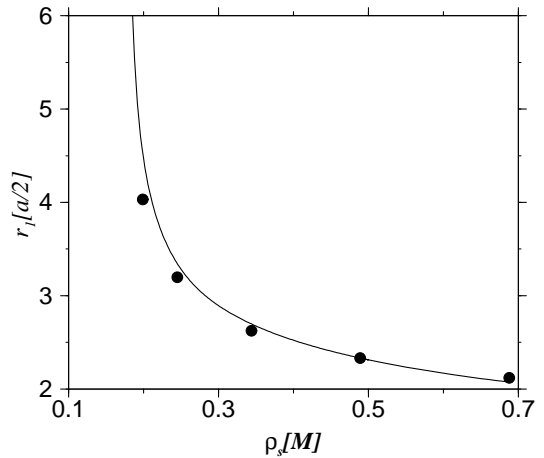}
  \caption{Radius $r_1$ from Eqn.~(\ref{eq:overcharging_radius}) at
    which overcharging sets in, for a DNA-sized and -charged cylinder
    ($r_0=7.86 \text {\AA}$ and $\sigma=0.190$ \Cmm), as a function
    of salt concentration $\rho_\roms$.  The distance $r_1$ is
    measured with respect to the cylinder surface in units of the
    ionic radius.  The dots are the results of molecular dynamics
    simulations while the solid line is the prediction of
    HNC/MSA.}
  \label{fig:DNA_r1_rho}
\end{figure}



In all cases, the HNC/MSA qualitative agreement with MD is very good. The success of HNC/MSA
theory indicates that local ion-size correlations are responsible for both
phenomena.  {\ch As it is concluded by Lozada-Cassou and Jim\'enez-\'Angeles
  in ref.~\cite{jimenez00a},  overcharging increases with an increase of
  the system  excluded volume: i.e., higher ionic size and/or concentration.
  If size correlations are not considered (as in PB
  theory) or are negligible, overcharging does not occur, even for divalent
  ions. On the other hand, overcharging can occur in a high excluded volume
  system {\ch even for monovalent ions}.  To illustrate this fact we have
  simulated two systems, for which the distance of closest approach to the
  charged rod is in both cases $9.98$ \AA, hence DNA-like. The first rod
  system has a charge parameter of $\xi = 4.2$ ($\sigma = 0.240$ \Cmm) and is
  immersed in a 1.0 Molar 1:1 electrolyte solution with ions of diameter
  $a=7.4$ \AA. The second rod system has an even higher charge parameter of
  $\xi = 10.5$ ( $\sigma = 0.40$ \Cmm) and is immersed in a 0.5 Molar 2:2
  electrolyte, but, this time, the diameter is only $a=1.0$ \AA.  The results are shown in
  Fig.~ \ref{fig:energy-entropy}, where the charge distribution function is
  plotted. Contrary to the general believe, the system with the large monovalent ions shows overcharging, both
  in the HNC/MSA and the MD results, and the location of the peaks coincide,
  whereas we see small differences in the height. Further, strikingly, the
  system with small divalent ions shows no sign of overcharging, although the
  charge parameter for this system is even higher. The shoulder in the
  distribution of the MD simulation indicates that the second ionic layer is
  almost neutral. A possible explanation is the following: The small ions have
  an energy of attraction of almost $30 k_\romB T$, hence the salt is actually
  present in the form of little overall neutral salt clusters. This is
  supported by visual inspection of the MD configurations.  These mostly
  neutral clusters are polarizable and hence tend to accumulate in the large
  field gradient surrounding the rod. The free counterions left can come only
  from the rods charge and can in no way lead to an overcharging.
A key prerequisite for this scenario is that the 
interaction energy between ions exceeds the interaction between 
ions and the rod.
Also consistent is, that the HNC/MSA calculation as well shows the 
vanishing of the overcharging.}

Finally, one might ask the question of how the effect of overcharging
depends on salt concentration. Obviously there can not be any
overcharging in the absence of salt, since then even a complete
``condensation'' of all ions would merely neutralize the rod. However,
it is not clear from the beginning whether an addition of just {\em
some\/} salt will immediately lead to overcharging.
Figure~\ref{fig:DNA_Pr_MEP_rho} shows HNC/MSA and MD distribution
functions $P(r)$ and mean electrostatic potentials $\psi(r)$ for a
DNA-sized {\md and} charged rod, immersed into a 2:2 electrolyte, for
{\md varying} salt concentrations. The phenomenon of overcharging is
indeed observed, but only at sufficiently high salt
concentration. These results are consistent with the fact that higher
electrolyte concentration increases the excluded volume, thus increases
overcharging ~\cite{jimenez00a}. Therefore, as proposed in Ref. ~\cite{jimenez00a},
higher excluded volume implies higher overcharging.  This may be also illustrated by defining $r_1$ to be the
radius at which overcharging sets in, i.e.,
\begin{equation}
  r_1 \; = \min\,\big\{r : \, P(r)=1\big\}.
  \label{eq:overcharging_radius}
\end{equation}
Fig.~\ref{fig:DNA_r1_rho} illustrates the measured $r_1$ together with
a hypernetted-chain prediction as a function of salt concentration
$\rho_\roms$.  Clearly, $r_1$ must increase with decreasing
$\rho_\roms$, since overcharging is reduced and consequently the size
of the charge-compensating layer must increase.  In fact,
hypernetted-chain theory predicts that $r_1(\rho_\roms)$ diverges at
some {\em finite\/} density $\rho_\roms^\infty\approx0.18 \,
\text{mol/l}$ corresponding to a salt Debye length of
$3.59\,\text{\AA}$, or roughly 200 salt molecules within the
simulation box. 
%
%
For the lower concentrations, the simulated $r_1$ values lie below the
hypernetted-chain prediction, but this is not a feature generally to
be expected. In Fig.~\ref{fig:DNA_Pr_MEP_rho} it can be seen that the
stronger overcharging in hypernetted-chain theory should normally lead
to a value of $r_1$ smaller than in the simulation.  {\md But} with
decreasing density the finite radius of the simulation cell becomes
relevant. {\md Indeed, zero-salt distribution functions reach the
value $1$ at the cell boundary, and not at infinity.}  A reduction of
the amount of added salt at fixed cell radius must necessarily lead to
values of $r_1$ smaller than the hypernetted chain prediction for the
bulk, since within a finite cell $r_1$ cannot diverge.  On the other
hand, for technical reasons, the MD cell radius can not be increased
indefinitely.  In any case, our MD results have not been able to
unambiguously detect overcharging at densities equal to or lower than
$0.2\,\text{M}$, {\md which is} in excellent agreement with the
HNC/MSA prediction {\md for the minimum amount of salt needed to
induce overcharging}. From Figs.~(\ref{fig:DNA_Pr_MEP_rho}) and (\ref{fig:DNA_r1_rho})
 it is clear that excluded volume effects
determine the occurrence of overcharging, as proposed by some of us \cite{jimenez00a}.


\section{Conclusion}

Theoretical and numerical studies of stiff linear polyelectrolytes,
immersed into a 2:2 RPM electrolyte have been presented.  The
qualitative agreement between the MD and HNC/MSA results is excellent.
For the particular case of DNA parameters, there is also a very good
quantitative agreement. {\md We argued that the better agreement in
the planar and spherical case reported in Refs.
\cite{degreve93,lozada98,lozada82} is most probably due to an
implementation of the short range interactions that is identical to
the theoretical model and not merely very close.}
%
{\md At low salt concentrations the difference between a finite cell used
in simulations and the $R=\infty$ system used in the HNC/MSA
calculations becomes important. Also, the macroion and ions electrical
fields are much less screened and, hence, the charge correlations
become more important \cite{lozada90}. Both these points are
responsible for the disagreement between theory and simulations at low
salt. While the former can in principle be resolved by increasing the
size of the simulation cell, the latter can be addressed by the three
point extension HNC/MSA theory \cite{lozada83a,lozada92}, which gives
a better account of these correlations \cite{jimenez00a}.}
%

In the past the HNC/MSA theory for an electrolyte next to a charged
rod has been derived \cite{lozada83a} and applied \cite{gonzales85a}
to simple DNA models.  In this paper we have shown this theory to be
qualitatively correct. In particular, for DNA parameters the
quantitative agreement is excellent.  To the best of our knowledge, in
Gonzales-Tovar \etal \cite{gonzales85a} overcharging has first
been predicted {\md theoretically}.  It was point out that this should
{\md entail an} electrophoresis mobility reversal.  In the PB theory
the $\zeta$-potential as a function of the cylinder charge density is
a monotonic function, {\md while in the HNC/MSA theory it is not}.  In
this paper we have shown the HNC/MSA prediction to be in agreement
with MD calculations. This non-monotonic behavior of the
$\zeta$-potential {\md has} important implications in electrophoresis
calculations. {\md The standard electrophoresis theory
\cite{wiersema66,obrien78} is based on the PB prediction for $\zeta$
and, hence, important differences should be found if the HNC/MSA
theory is applied to the electrophoresis problem.}

Recently, Lozada-Cassou \etal \cite{lozada99} extended the
electrophoresis theory to include ionic size effects, through the
HNC/MSA theory. It was applied to spherical macroions.  A mobility
reversal and a very non-linear behavior of the mobility, as a function
of the $\zeta$-potential, were predicted. These predictions were found
to be in agreement with experimental results.  Contrary to the
standard electrophoresis theory, {\md the mobility was shown to be
non-universal.}
This is due, precisely, to the non-monotonic behavior of the
$\zeta$-potential.  The HNC/MSA and MC simulations for spherical and
planar macroions do not show {\md a maximum in $\zeta(\sigma)$ as
pronounced as that reported here for the cylindrical geometry}.
Therefore, {\ch we believe that our HNC/MSA and MD results are}
particularly relevant.

In view of the qualitative agreement between HNC/MSA and MD and the
quantitative and qualitative disagreement of these with Poisson-Boltzmann
theory, we conclude that for concentrated and divalent electrolytes,
influenced by charged polyelectrolytes, the size correlations must be taken
{\ch properly} into account to describe the cylindrical double layer. Our MD
results clearly indicate that overcharging strongly depends on the system 
excluded volume, as proposed by some of us \cite{jimenez00a}.  While maximum 
overcharging does not seem to have a limit, the minimum conditions to have overcharging
depend on many different ways in which ionic charge, size and concentration, 
and surface charge density participate in a system, as we showed in Figs. 
(\ref{fig:DNA_Pr_MEP_lambda}), (\ref{fig:energy-entropy}) and (\ref{fig:DNA_Pr_MEP_rho}) 

\section{Acknowledgments}

CH and MLC wish to thank W. Gelbart and The Institute for Theoretical
Physics, University of California at Santa Barbara, where this project
was started, for their hospitality.  MLC which to thank C. Holm and
K. Kremer for their hospitality at Mainz.  FJA and MLC gratefully
acknowledge the financial support of INDUSTRIAS NEGROMEX.  CH and MD
acknowledge a large computer time grant hkf06 from NIC J\"ulich and
financial support by the German Science foundation.











\end{document}